\documentclass[12pt]{article}

\usepackage{amsmath}
\usepackage{amssymb}

\begin{document}

\title{Curvature Couplings in
$\mathcal{N}=(2,2)$ Nonlinear Sigma Models on $S^2$}

\author{Bei Jia, Eric Sharpe\\
        \footnotesize{Department of Physics, Virginia Tech,
Blacksburg, VA 24061 USA}\\
\footnotesize{{\tt beijia@vt.edu}, {\tt ersharpe@vt.edu}}}

%\date{}

\maketitle

\begin{abstract}
Following recent work on GLSM localization,
we work out curvature couplings for rigidly supersymmetric
nonlinear sigma models with superpotential for general target spaces,
describing both ordinary and
twisted chiral superfields on round two-sphere worldsheets.  We briefly discuss
why, unlike four-dimensional theories, there are no constraints on
Kahler forms in these theories.
We also briefly discuss
general issues in topological twists of such theories.
\end{abstract}

\newpage

\section{Introduction}

In recent years, curvature couplings in
rigidly supersymmetric nonlinear sigma models on
nontrivial spacetime manifolds of various dimensions have been discussed by
several groups, see
\cite{allanetal, fest-seib, butter-kuzenko, ours, samtleben-tsimpis,
liu-zayas-reichmann, dumitrescu-festuccia-seiberg, dumitrescu-festuccia,
kehagias-russo, closset-dumitrescu-gestuccia-komargodski,
samtleben-sezgin-tsimpis,ktz1,ckmtz,htz1} and references contained therein.
Often, supersymmetry on the nontrivial spacetime manifold will constrain
the target space in some fashion, by {\it e.g.} requiring the
K\"ahler form on the target space to be cohomologically trivial.
Furthermore,
supersymmetric localization techniques have been applied to
theories of this form
to obtain
quantum mechanically exact results such as the partition function
\cite{pestun, kapusin-willett-yaakov, jafferis, hama-hosomichi-lee,
hosomichi-seong-terashima, benini-cremonesi, doroud-gomis-le floch-lee}.

Recently these methods were applied
to two-dimensional
gauged linear sigma models (GLSMs) to derive exact expressions for
partition functions \cite{benini-cremonesi,doroud-gomis-le floch-lee},
which has quickly led to some
interesting new computational methods and results for Gromov-Witten invariants
\cite{jockers-kumar-lapan-morrison-romo,ncgw,hkm}
and the Seiberg-Witten K\"{a}hler potential \cite{park-song},
as well as other insights into older results \cite{gomis-lee}.
As part of that work,
the papers \cite{benini-cremonesi,doroud-gomis-le floch-lee}
worked out curvature couplings in two-dimensional linear sigma
models whose target spaces are vector spaces.

In this paper, we return to \cite{benini-cremonesi,doroud-gomis-le floch-lee}
to re-examine general rigidly supersymmetric
nonlinear sigma models with potential and work out
curvature couplings,
for more general target spaces (and with $U(1)_R$ actions described by
more general holomorphic Killing vectors), for both
ordinary and twisted chiral supermultiplets, on constant-curvature (round)
two-sphere worldsheets, so as to give some insight into the rather complicated
linear actions of \cite{benini-cremonesi,doroud-gomis-le floch-lee}.

We begin in section~\ref{sect:general-action} by working out general
nonlinear sigma models with potential for ordinary chiral supermultiplets
on round two-spheres.  To add a superpotential in a theory
of ordinary chiral multiplets, one must extend the flat-space $U(1)_R$
symmetry by a holomorphic Killing vector, which generates {\it e.g.}
a curvature-dependent potential term in the action.
We also discuss why two-dimensional theories of this sort do not have
constraints on the Kahler form on the target space, unlike typical behavior
in four dimensional theories.

In section~\ref{sect:twistedchiral}, we perform analogous analyses
for twisted chiral multiplets.
Here, the curvature couplings have
a different form than for ordinary chiral multiplets.
For example, although one can extend the
$U(1)_R$ of the twisted chiral theory by a holomorphic Killing vector
field, that same field can be re-absorbed into the auxiliary field of the
multiplet, and so its presence is optional.  Moreover, these same couplings
naively break a duality between the flat-worldsheet chiral and twisted
chiral theories.

We conclude in
section~\ref{sect:tft} with a discussion of topological twists in the
presence of such curvature couplings.
An appendix further discussions the relationship between ordinary and twisted
chiral multiplets, and their topological twists.

\section{Ordinary chiral supermultiplets}
\label{sect:general-action}

In this section, we will discuss curvature couplings for ordinary
chiral multiplets on an $S^2$ with a constant curvature metric,
the `round' $S^2$.

The rigid $\mathcal{N}=(2,2)$ supersymmetry algebra on $S^2$ with
Euclidean signature is
\cite{benini-cremonesi, doroud-gomis-le floch-lee}
\begin{displaymath}
OSp(2|2) \: \cong \: SU(2|1),
\end{displaymath}
whose bosonic subalgebra is $SU(2)\times U(1)_R$.
The $SU(2)$ factor represents the isometries of $S^2$,
while the $U(1)_R$ factor is the vector-like R-symmetry,
which is now contained in the supersymmetry algebra
rather than being an outer automorphism of it.

Our spinor notation will follow
\cite{doroud-gomis-le floch-lee}[p. 47], \cite{wess-bagger}.
Spinors are multiplied as
\begin{displaymath}
\lambda\psi \: = \: \lambda^{\alpha}\varepsilon_{\alpha\beta}\psi^{\beta}
\: = \: \psi\lambda,
\end{displaymath}
where $\varepsilon_{21}=-\varepsilon_{12}=1$,
$\varepsilon_{11}=-\varepsilon_{22}=0$. The two-dimensional $\gamma_m$
matrices are given by the Pauli matrices in local frame coordinates:
$\gamma_m=\sigma_m$, where
\begin{displaymath}
\sigma_1=\begin{pmatrix}
                                                                                                                          0 & 1 \\
                                                                                                                          1 & 0 \\
                                                                                                                        \end{pmatrix}
, \: \: \sigma_2=\begin{pmatrix}
            0 & -i \\
            i & 0 \\
          \end{pmatrix},
\end{displaymath}
with
\begin{displaymath}
\gamma_3=-i\sigma_1\sigma_2=\begin{pmatrix}
                                       1 & 0 \\
                                       0 & -1 \\
                                     \end{pmatrix}.
\end{displaymath}

With our notation, the explicit form of the supersymmetry algebra
$SU(2|1)$ is \cite{doroud-gomis-le floch-lee}:
\begin{equation} \label{susyalgebra}
\begin{split}
\{Q_{\alpha}, \bar{Q}_{\beta}\} &= \gamma_{\alpha\beta}^m J_m-\frac{1}{2}\varepsilon_{\alpha\beta}R \quad
[R, Q_{\alpha}]=Q_{\alpha} \qquad \qquad \ [R, \bar{Q}_{\alpha}]=-\bar{Q}_{\alpha}\\
[J_m,J_n]&=i\varepsilon_{mnl}J_l \qquad \qquad \ [J_m, Q_{\alpha}]=-\frac{1}{2}\gamma_{m\alpha}^{\beta}Q_{\beta} \quad
[J_m, Q\bar{}_{\alpha}]=-\frac{1}{2}\gamma_{m\alpha}^{\beta}\bar{Q}_{\beta}
\end{split}
\end{equation}
where $Q_{\alpha}$ and $\bar{Q}_{\alpha}$ are the supersymmetry generators,
$J_m$ are the generators of the $SU(2)$ isometry of $S^2$,
and $R$ is the generator of the $U(1)_R$ symmetry.

The space of Killing spinors on $S^2$ is four-dimensional \cite{killing};
a useful basis of this space consists of two Killing spinors called
positive Killing spinors and two Killing spinors called negative Killing
spinors \cite{benini-cremonesi}. Let's denote the positive Killing spinors as $\zeta,\bar{\zeta}$, which are independent from each other and satisfy the same
Killing spinor equation
\begin{equation} \label{killingfull}
\begin{split}
\nabla_{m}\zeta-\frac{i}{2r}\gamma_m\zeta  &=0,\\
\nabla_{m}\bar{\zeta}-\frac{i}{2r}\gamma_m\bar{\zeta}  &=0.
\end{split}
\end{equation}

Some useful spinor identities include:
\begin{displaymath}
\lambda \psi \: = \: + \psi \lambda, \: \: \:
\lambda\gamma_m\psi=\lambda^{\alpha}(\gamma_m)_{\alpha}^{\beta}\psi_{\beta}
\: = \:
-\psi\gamma_m\lambda,
\: \: \:
\gamma_m\gamma_n=g_{mn}+i\varepsilon_{mn}\gamma_3,
\: \: \:
\gamma^m \gamma^n \gamma_m = 0,
\end{displaymath}
\begin{displaymath}
( \psi_1 \psi_2) \psi_3 \: + \: (\psi_2 \psi_3) \psi_1 \: + \:
(\psi_3 \psi_1) \psi_2 \: = \: 0.
\end{displaymath}

An Euclidean $\mathcal{N}=(2,2)$ chiral multiplet in
two dimensions contains components
\begin{displaymath}
(\phi^i,\bar{\phi}^{\bar{\imath}},\psi^i,\bar{\psi}^{\bar{\imath}},
F^i,\bar{F}^{\bar{\imath}}).
\end{displaymath}
The chiral fields $\phi^i,\bar{\phi}^{\bar{\imath}}$ parametrize the
target space $M$, which is a K\"{a}hler manifold.
As observed in \cite{benini-cremonesi, doroud-gomis-le floch-lee},
the $U(1)_R$ charges of the chiral fields $\phi^i$ enter the definition of
the Lagrangian of the linear sigma models on $S^2$.
To construct nonlinear sigma models on a worldsheet
$S^2$, one needs to use the
holomorphic Killing vector $X=X^i\partial_i$
corresponding to the $U(1)_R$ symmetry, which should be interpreted as
an isometry on the target space $M$ of the nonlinear sigma model.

The general Lagrangian governing $\mathcal{N}=(2,2)$
nonlinear sigma models on $S^2$ is
\begin{equation}  \label{nlsm}
\begin{split}
\mathcal{L} \: = \:
& g_{i\bar{\jmath}}\partial_{m}\phi^i\partial^{m}\bar{\phi}^{
\bar{\jmath}} \: - \: ig_{i\bar{\jmath}}\bar{\psi}^{\bar{\jmath}}\gamma^{m}
\nabla_{m}\psi^i \: + \: g_{i\bar{\jmath}}F^i
\bar{F}^{\bar{\jmath}} \: - \: F^i(\frac{1}{2}g_{i\bar{\jmath},\bar{k}}
\bar{\psi}^{\bar{\jmath}}\bar{\psi}^{\bar{k}}-W_i)\\
 & - \: \bar{F}^{\bar{\imath}}(\frac{1}{2}g_{j\bar{\imath},k}\psi^j\psi^k-
\bar{W}_{\bar{\imath}}) \: - \: \frac{1}{2}W_{ij}\psi^i\psi^j
\: - \: \frac{1}{2}\bar{W}_{\bar{\imath}\bar{\jmath}}\bar{\psi}^{\bar{\imath}}
\bar{\psi}^{\bar{\jmath}} \: + \:
\frac{1}{4}g_{i\bar{\jmath},k\bar{\ell}}(\psi^i\psi^k)
(\bar{\psi}^{\bar{\jmath}}\bar{\psi}^{\bar{\ell}})\\
&- \: \frac{1}{4r^2}g_{i\bar{\jmath}}X^iX^{\bar{\jmath}} \: + \:
\frac{i}{4r^2}K_iX^i \: - \: \frac{i}{4r^2}K_{\bar{\imath}}X^{\bar{\imath}}
\: - \: \frac{i}{2r}g_{i\bar{\jmath}}\bar{\psi}^{\bar{\jmath}} \psi^j
\nabla_jX^i,
\end{split}
\end{equation}
where\footnote{
As an aside, terms closely analogous to the
curvature-dependent terms in the Lagrangian above
have been discussed in two-dimensional theories in a different context
in \cite{gates84}.
} $r$ is the radius of $S^2$, $K$ is the K\"{a}hler potential
of the target space $M$, $W$ is the superpotential, and
\begin{equation}
\begin{split}
%\mathcal{D}_m\phi^i &=\nabla_m\phi^i,\\
\nabla_{m}\psi^i &= \widetilde{\nabla}_m\psi^i+\Gamma^i_{jk}(\partial_m\phi^j)
\psi^k,\\
\nabla_j X^i &= \partial_j X^i+\Gamma^i_{jk}X^k,
\end{split}
\end{equation}
with $\Gamma^i_{jk}$ the Christoffel symbols on the target space $M$,
and $\widetilde{\nabla}_m$ denotes the pure worldsheet spin connection
covariant derivative.
Integrating out the auxiliary fields yields
\begin{equation}
\begin{split}
\mathcal{L} \: = \: &
 g_{i\bar{\jmath}}\partial_{m}\phi^i\partial^{m}\bar{\phi}^{
\bar{\jmath}} \: - \: ig_{i\bar{\jmath}}\bar{\psi}^{\bar{\jmath}}\gamma^{m}
\nabla_{m}\psi^i \: + \:
R_{i \bar{\jmath} k \bar{\ell}} (\psi^i\psi^k)
(\bar{\psi}^{\bar{\jmath}}\bar{\psi}^{\bar{\ell}})\\
& \: - \:
g^{i \bar{\jmath}} W_i \bar{W}_{\bar{\jmath}}
\: - \: \frac{1}{2} \nabla_i \partial_j W \psi^i \psi^j
\: - \: \frac{1}{2} \nabla_{\bar{\imath}} \partial_{\bar{\jmath}}
\bar{W} \bar{\psi}^{\bar{\imath}} \bar{\psi}^{\bar{\jmath}} \\
&\: - \: \frac{1}{4r^2}g_{i\bar{\jmath}}X^iX^{\bar{\jmath}} \: + \:
\frac{i}{4r^2}K_iX^i \: - \: \frac{i}{4r^2}K_{\bar{\imath}}X^{\bar{\imath}}
\: - \: \frac{i}{2r}g_{i\bar{\jmath}}\bar{\psi}^{\bar{\jmath}} \psi^j
\nabla_jX^i.
\end{split}
\end{equation}
This Lagrangian reduces the the usual $\mathcal{N}=(2,2)$ nonlinear sigma model on flat two-dimensional space when $r\rightarrow \infty$, as one would expect.

There are two conditions on the data above.
First, the K\"ahler potential is invariant under the isometry defined
by the holomorphic Killing vector $X$, which means that on each
coordinate patch,
\begin{equation} \label{constraint1}
{\cal L}_X K \: = \: X^i K_i \: + \: X^{\overline{\imath}} K_{\overline{\imath}}
\: = \: 0
\end{equation}
and across coordinate patches, $K\mapsto K+f(\phi)+\bar{f}(\bar{\phi})$,
the $f$'s are constrained to obey
\begin{equation}
\sum_i X^i\partial_i f(\phi) \: = \: 0.
\end{equation}
On a K\"ahler manifold with a holomorphic isometry $X$, for data on
a good open cover, one can always choose $K$'s, $f$'s to obey these
constraints \cite{bryant-priv}.

The second condition says that the superpotential $W$ is homogeneous
of degree 2 under the vector-like $U(1)_R$ symmetry, meaning
\begin{equation}  \label{constraint2}
2 W  \: - \: i X^i W_i  \: =  \: 0
\end{equation}
(up to an additive constant).
In particular, if $X = 0$, then the superpotential must vanish (up to a
constant).

The Lagrangian above is invariant under the following supersymmetry
transformations
\begin{equation} \label{rigid-chiral-trans}
\begin{split}
\delta \phi^i &= \zeta\psi^i,\\
\delta \bar{\phi}^{\bar{\imath}} &= \bar{\psi}^{\bar{\imath}}\bar{\zeta},\\
\delta \psi^i &= i\gamma^m\bar{\zeta}\partial_m \phi^i -
\frac{i}{2r}\bar{\zeta}X^i +\zeta F^i,\\
\delta \bar{\psi}^{\bar{\imath}} &= i\gamma^m\zeta\partial_m
\bar{\phi}^{\bar{\imath}}
+ \frac{i}{2r}\zeta X^{\bar{\imath}} +\bar{\zeta} \bar{F}^{\bar{\imath}},\\
\delta F^i &= i\bar{\zeta}\gamma^m \tilde{\nabla}_m\psi^i
+\frac{i}{2r}\bar{\zeta}\psi^j \partial_j X^i,\\
\delta \bar{F}^{\bar{\imath}} &= i\zeta\gamma^m
\tilde{\nabla}_m\bar{\psi}^{\bar{\imath}}
-\frac{i}{2r}\zeta\bar{\psi}^{\bar{\jmath}} \partial_{\bar{\jmath}}
X^{\bar{\imath}}.
\end{split}
\end{equation}
(up to total derivatives) provided the
Killing spinor equations (\ref{killingfull}) are satisfied,
together with the constraints~(\ref{constraint1}), (\ref{constraint2}).

In the special case that $X=0$ (and hence $W=0$), the Lagrangian and
supersymmetry transformations are identical to those on flat space.
One can show
that the flat-space Lagrangian is invariant under supersymmetry
transformations defined by a Killing spinor appropriate for $S^2$,
not just a constant spinor.  Thus, the lagrangian with $X=0$ is consistent
with supersymmetry on both $S^2$ and ${\mathbb R}^2$, as one would expect.

In the linear case, {\it i.e.} when the target space $M=\mathbb{C}$
with the trivial K\"{a}hler potential $K=\bar{\phi}\phi$ and the
$U(1)_R$ holomorphic Killing vector $X=-iq\phi\frac{d}{d\phi}$,
the Lagrangian~(\ref{nlsm}) reduces to the Lagrangian of the chiral
multiplet in \cite{benini-cremonesi, doroud-gomis-le floch-lee}
(where $q$ is the $U(1)_R$ charge of the chiral field $\phi$). The full
gauged linear sigma model in \cite{benini-cremonesi, doroud-gomis-le floch-lee}
can also be obtained by applying a decoupling gravity procedure
analogous to the one in \cite{allanetal, fest-seib} to the coupled theory of (2,2)
gauged linear sigma model and (1,1) supergravity model in
\cite{nishino}.

From the supersymmetry transformations above, we can get some insight
into the constraint on $W$.  Specifically, note that with the Killing
spinor condition, the supersymmetry variation of the auxiliary field
can be written
\begin{displaymath}
\delta F^i \: = \: i \bar{\zeta} \gamma^m \tilde{\nabla}_m \psi^i \: - \:
\frac{1}{2} (\nabla_m \bar{\zeta}) \gamma^m \psi^j \partial_j X^i
\end{displaymath}
If $\partial_j X^i = - 2 i \delta_j^i$, then $\delta F^i$ is a total derivative.
For example, in the linear case above, this is the statement that
when $q=2$, $\delta F$ is a total derivative.
Ultimately that factor of two is
the reason why supersymmetry requires that $W$ be
homogeneous of degree two under the action of $X$.

Now, let us describe the vector $U(1)_R$ symmetry explicitly.
Infinitesimally, the action of this global $U(1)_R$ symmetry on the fields is
given by
\begin{equation} \label{u1}
\begin{split}
\delta \phi^i & =  \epsilon X^i, \\
\delta \phi^{\overline{\imath}} & =  \epsilon X^{\overline{\imath}}, \\
\delta \psi^i & =  \epsilon\left( \partial_j X^i \: - \: i
\delta^i_j \right) \psi^j, \\
\delta \psi^{\overline{\imath}} & =  \epsilon\left(
\partial_{\overline{\jmath}} X^{\overline{\imath}} \: + \:
i  \delta^{\overline{\imath}}_{\overline{\jmath}} \right)
\psi^{\overline{\jmath}}, \\
\delta F^i & =  \epsilon\left( \partial_j X^i \: - \: 2 i
\delta^i_j\right) F^j - \frac{\epsilon}{2}\partial_k\partial_j X^i \psi^j\psi^k, \\
\delta F^{\overline{\imath}} & = \epsilon\left(
\partial_{\overline{\jmath}} X^{\overline{\imath}} \: + \: 2 i
\delta^{\overline{\imath}}_{\overline{\jmath}} \right) F^{\overline{\jmath}}- \frac{\epsilon}{2}\partial_{\bar{k}}\partial_{\bar{\jmath}}
X^{\bar{\imath}} \psi^{\bar{\jmath}}\psi^{\bar{k}}
\end{split}
\end{equation}
where $\epsilon$ is a real constant parametrizing the global $U(1)_R$. Our Lagrangian (\ref{nlsm}) is invariant under this symmetry.

We should observe that even for (2,2) supersymmetric
theories on ${\mathbb R}^2$ instead of
$S^2$, the $U(1)_R$ symmetry sometimes involves an action on bosons,
and hence involves a holomorphic Killing vector field $X$.
In fact, the explicit transformations (\ref{u1}) also applies to the usual
$\mathcal{N}=(2,2)$ nonlinear sigma models on $\mathbb{R}^2$.
In general, a global symmetry of a nonlinear sigma model on any spacetime
should act on the bosonic fields as a Killing vector on the target space.
We believe that (\ref{u1}) should hold for any two-dimensional nonlinear sigma
models with a vector $U(1)_R$ symmetry, regardless of the two-dimensional
spacetime they are defined on.

In the case of four-dimensional rigidly supersymmetric theories on
spacetimes such as $S^4$ and AdS$_4$, supersymmetry imposes constraints
on the theory (see {\it e.g.} \cite{allanetal,fest-seib,butter-kuzenko,ours}),
such as a constraint that the K\"ahler form on the target space be
cohomologically-trivial.  In two dimensional theories, on the other
hand, we have found no analogous constraint.

Mechanically, one way to understand this lack of constraints on
two-dimensional theories is to think of a two-dimensional
theory as a dimensional reduction of a four-dimensional theory on
${\bf R} \times \Sigma_3$ (for $\Sigma_3$ a three-manifold), or
${\bf R}^2 \times \Sigma_2$ (for $\Sigma_2$ a two-manifold).  Such four
dimensional theories were unconstrained by supersymmetry; constraints
only existed in four dimensions when all four spacetime
directions were `wrapped up'
nontrivially in the topology, when none were flat.
Another more abstract way to think about this in the context
of the decoupling procedure is as follows\footnote{
We would like to thank I.~Melnikov for making this observation.}.
In four dimensional supergravities,
the Fayet-Iliopoulos parameter\footnote{See
\cite{nati0,dist-shar,banks-seib,hs} for a recent
discussion of the Fayet-Iliopoulos parameter in supergravity, and how
old obstruction issues summarized in {\it e.g.} \cite{dienes-thomas} can be
circumvented.},
the curvature of the Bagger-Witten line bundle, and so forth
are weighted by inverse factors of the four-dimensional Planck mass.
The decoupling limit of \cite{allanetal,fest-seib} involves
sending the Planck mass to infinity, which necessarily truncates those
terms, and leaves one with a rigidly supersymmetric theory in which
Fayet-Iliopoulos parameters vanish and K\"{a}hler forms are exact.
By contrast, in two dimensions, the ``Planck mass'' is dimensionless.
Hence, the procedure of decoupling gravity in two dimensions is a formal way
of obtaining rigid supersymmetric theories from supergravity theories,
with no further constraints on the target space geometry.
Thus, one should not be surprised to find no constraints on the target
space geometry in two dimensional cases.

\section{Twisted chiral supermultiplets}  \label{sect:twistedchiral}

In two dimensions, there is another $\mathcal{N}=(2,2)$
supermultiplet known as the twisted chiral multiplet.
The field content of a twisted chiral multiplet is the same as
that of an ordinary chiral multiplet:
\begin{equation*}
(\rho,\bar{\rho},\chi,\bar{\chi},G,\bar{G}).
\end{equation*}
The fields $\rho^i,\bar{\rho}^{\bar{\imath}}$ are bosons describing maps
into a target space $\tilde{M}$, which is required\footnote{
We do not attempt to consider $H$-flux backgrounds in this paper.
} to be a
K\"{a}hler manifold.  The fields $G$, $\bar{G}$ are auxiliary fields.

Although the flat-worldsheet action of a twisted chiral multiplet is
identical to that of an ordinary chiral multiplet, the curvature couplings
to a superpotential are of a very different form.

The fields
$\chi,\bar{\chi}$ are Dirac spinors.  In a twisted chiral multiplet,
their components mix holomorphic and antiholomorphic target space
indices:
\begin{equation}
\chi=\left(
                                 \begin{array}{c}
                                   \chi^i_-  \\
                                   \chi^{\bar{\imath}}_+\\
                                 \end{array}
                               \right),
\quad
\bar{\chi}=\left(
                                 \begin{array}{c}
                                   \bar{\chi}^{\bar{\imath}}_- \\
                                   \bar{\chi}^i_+ \\
                                 \end{array}
                               \right)
\end{equation}

In this section, we will work with Killing spinors $\zeta$, $\bar{\zeta}$
obeying
\begin{equation}  \label{eq:killing-twisted}
\begin{split}
\nabla_m \zeta & = \frac{i}{2r} \gamma_m \zeta, \\
\nabla_m \bar{\zeta} & = - \frac{i}{2r} \gamma_m \bar{\zeta},
\end{split}
\end{equation}
a slightly different convention than we used for ordinary chiral multiplets.

Although the flat-worldsheet action of a twisted chiral multiplet is
identical to that of an ordinary chiral multiplet, the curvature couplings
to a superpotential are of a very different form.
In the Killing spinor convention~(\ref{eq:killing-twisted}),
the most general $\mathcal{N}=(2,2)$ Lagrangian for twisted chiral multiplets
on a round $S^2$ is
\begin{equation}  \label{twisted-lagrangian}
\begin{split}
\mathcal{L}_T \: = \:
& g_{i\bar{\jmath}}\partial_{m}\rho^i\partial^{m}\bar{\rho}^{
\bar{\jmath}} \:  + \:
2ig_{i\bar{\jmath}}\bar{\chi}^{\bar{\jmath}}_{-}\nabla_{\bar{z}}\chi^i_{-}
\: + \: 2ig_{i\bar{\jmath}}\chi^{\bar{\jmath}}_{+}\nabla_{z}\bar{\chi}^i_{+}
\: +  \: g_{i\bar{\jmath}}G^i\bar{G}^{\bar{\jmath}}\\
&\: - \: G^i\left( i g_{i\bar{\jmath},\bar{k}} \bar{\chi}^{\bar{\jmath}}_-
\chi^{\bar{k}}_+ \: - \: \mathcal{W}_i\right) \: - \: i \mathcal{W}_{ij}
\chi^i_-\bar{\chi}^j_+ \\
& \: - \: \bar{G}^{\bar{\imath}}
 \left(i g_{\bar{\imath}j,k}\chi^j_-\bar{\chi}^k_+ \: - \:
\bar{\mathcal{W}}_{\bar{\imath}}\right)
\: - \: i \bar{\mathcal{W}}_{\bar{\imath}\bar{\jmath}}\bar{\chi}^{\bar{\imath}}_-\chi^{\bar{\jmath}}_+ \\
&\: + \: g_{i\bar{\jmath},k\bar{l}}\bar{\chi}_+^i
\chi_+^{\bar{\jmath}} \chi_-^k \bar{\chi}_-^{\bar{l}} \\
& \: + \: \frac{i}{r} \mathcal{W} \: - \: \frac{i}{r} \bar{\mathcal{W}}
,
\end{split}
\end{equation}
where, as in the case of ordinary chiral multiplets,
$g_{i\bar{\jmath}}$ is the K\"{a}hler metric on $\tilde{M}$,
and $\mathcal{W}$ is the twisted
superpotential.  (For notational simplicity, we have chosen to write the
spinors in the lagrangian above in chiral components.)
Notice that the twisted superpotential $\mathcal{W}$ is
coupled to the curvature of $S^2$, unlike the superpotential $W$ of the
ordinary chiral multiplets.  Furthermore, unlike ordinary chiral
multiplets, no holomorphic Killing vector is needed to define the
superpotential.  (It is straightforward to check that a
curvature coupling of this form
is incompatible with the supersymmetry of the ordinary
chiral multiplets.)  This Lagrangian generalizes\footnote{
We have absorbed the weight $\Delta$ in \cite{gomis-lee} in field
redefinitions; later in this section we shall give an alternative form
of the lagrangian in which that weight reappears, in terms of a
vector $Y$.}
that given in \cite{gomis-lee}[equ'ns (4.2), (4.5)] for flat target spaces.
It also reduces to the
flat $\mathbb{R}^2$ lagrangian when $r\rightarrow \infty$, as expected
(compare appendix~\ref{app:twistedchirals}).

The fermions couple to the following bundles:
\begin{displaymath}
\begin{array}{ll}
\bar{\chi}_+^i \: \in \: \Gamma_{C^{\infty}}\left( K_{\Sigma}^{1/2}
\otimes \rho^* T^{1,0} \tilde{M}
\right), &
\chi_-^i \: \in \: \Gamma_{C^{\infty}}\left(
\overline{K}_{\Sigma}^{1/2} \otimes \left( \rho^*  T^{0,1} \tilde{M}
\right)^* \right), \\
\chi_+^{\bar{\imath}} \: \in \: \Gamma_{C^{\infty}}\left(
K_{\Sigma}^{1/2} \otimes \left( \rho^* T^{1,0} \tilde{M} \right)^*
\right), &
\bar{\chi}_-^{\bar{\imath}} \: \in \: \Gamma_{C^{\infty}}\left(
\overline{K}_{\Sigma}^{1/2} \otimes \rho^* T^{0,1} \tilde{M} \right).
\end{array}
\end{displaymath}
where $\Sigma$ is the worldsheet (here, $S^2$), $K_{\Sigma}$ and
$\overline{K}_{\Sigma}$ are the holomorphic and antiholomorphic canonical
bundles.

The above Lagrangian is invariant under the following
supersymmetry transformations:
\begin{equation} \label{susy-twisted}
\begin{split}
\delta \rho^i &= i \bar{\zeta}_+\chi^i_- + i \zeta_-\bar{\chi}^i_+,\\
\delta \bar{\rho}^{\bar{\imath}} &= i \bar{\zeta}_- \chi^{\bar{\imath}}_+
 + i \zeta_+ \bar{\chi}^{\bar{\imath}}_-,\\
\delta \bar{\chi}^i_+ &= -2 \bar{\zeta}_-\bar{\partial}\rho^i
- \bar{\zeta}_+ G^i,\\
\delta \chi^i_- &= -2 \zeta_+\partial\rho^i +\zeta_- G^i,\\
\delta \chi^{\bar{\imath}}_+ &= -2 \zeta_-\bar{\partial}
\bar{\rho}^{\bar{\imath}} - \zeta_+ \bar{G}^{\bar{\imath}},\\
\delta \bar{\chi}^{\bar{\imath}}_- &= -2 \bar{\zeta}_+
\partial\bar{\rho}^{\bar{\imath}}  +\bar{\zeta}_- \bar{G}^{\bar{\imath}},\\
\delta G^i &= 2i(\zeta_+\tilde{\nabla}_z\bar{\chi}^i_+ - \bar{\zeta}_-
\tilde{\nabla}_{\bar{z}}\chi^i_-) ,\\
\delta \bar{G}^{\bar{\imath}} &= 2i(\bar{\zeta}_+
\tilde{\nabla}_z\chi^{\bar{\imath}}_+
- \zeta_-\tilde{\nabla}_{\bar{z}}\bar{\chi}^{\bar{\imath}}_-) ,
\end{split}
\end{equation}
provided that the Killing spinor equations~(\ref{eq:killing-twisted})
are satisfied.

In order to check supersymmetry, it is useful to write down the Killing
spinor equations~(\ref{eq:killing-twisted}) in chiral components:
\begin{displaymath}
\begin{array}{ll}
\nabla_z \zeta_- \: = \: 0, &
\nabla_{\bar{z}} \zeta_- \: = \: \frac{i}{2r} \zeta_+, \\
\nabla_z \zeta_+ \: = \: \frac{i}{2r} \zeta_-, &
\nabla_{\bar{z}} \zeta_+ \: = \: 0, \\
\: & \: \\
\nabla_z \bar{\zeta}_- \: = \: 0, &
\nabla_{\bar{z}} \bar{\zeta}_- \: = \: - \frac{i}{2r} \bar{\zeta}_+, \\
\nabla_z \bar{\zeta}_+ \: = \: - \frac{i}{2r} \bar{\zeta}_-, &
\nabla_{\bar{z}} \bar{\zeta}_+ \: = \: 0.
\end{array}
\end{displaymath}

Note that the supersymmetry transformation of the auxiliary field above
is not a total derivative,
even though the theory is supersymmetric and
contains a superpotential, which is very different from the behavior of
ordinary chiral multiplets.  We will shortly describe how one can couple a
holomorphic Killing vector field, which could be used to make $\delta G^i$
a total derivative, but unlike the case of ordinary chiral multiplets,
it is not necessary in order to couple a superpotential.  One suspects
that this may be linked to the existence of superfield representations on
spheres, but we will not speculate further in that direction.

For $\mathcal{N}=(2,2)$ nonlinear sigma models on $\mathbb{R}^2$,
the Lagrangian of twisted chiral multiplets
can be obtained by a simple ``twist''
from the Lagrangian of ordinary chiral multiplets, just by dualizing
the tangent bundle $TM$ to $T^*M$ on the right-movers, as elaborated
in appendix~\ref{app:twistedchirals}.
However, notice that for a nonzero superpotential,
the above Lagrangian for twisted chiral fields on $S^2$
cannot be obtained from the Lagrangian of chiral fields on $S^2$,
given in equation~(\ref{nlsm}), in a similar fashion,
because of different couplings to the curvature of $S^2$.

To compare to the lagrangian for twisted chiral multiplets given in
\cite{gomis-lee}, one performs a slight field redefinition.
Specifically, redefine $G^i$ to be $G^i + (i/r) Y^i$ for $Y$ a
vector field.  Then, the lagrangian becomes
\begin{equation}
\begin{split}
\mathcal{L}_T
\: = \:
& g_{i\bar{\jmath}}\partial_{m}\rho^i\partial^{m}\bar{\rho}^{
\bar{\jmath}} \: + \:
2ig_{i\bar{\jmath}}\bar{\chi}^{\bar{\jmath}}_{-}\nabla_{\bar{z}}\chi^i_{-}
\: + \: 2ig_{i\bar{\jmath}}\chi^{\bar{\jmath}}_{+}\nabla_{z}\bar{\chi}^i_{+}
\: + \: g_{i\bar{\jmath}}\left( G^i + \frac{i}{r} Y^i \right)
\left( \bar{G}^{\bar{\jmath}} - \frac{i}{r} Y^{\bar{\jmath}} \right)\\
& \: - \: \left( G^i + \frac{i}{r} Y^i \right)
\left(i g_{i\bar{\jmath},\bar{k}} \bar{\chi}^{\bar{\jmath}}_-
\chi^{\bar{k}}_+ \: - \: \mathcal{W}_i\right) \: - \: i \mathcal{W}_{ij}
\chi^i_-\bar{\chi}^j_+ \\
& \: - \: \left( \bar{G}^{\bar{\imath}} - \frac{i}{r} Y^{\bar{\imath}} \right)
 \left(i g_{\bar{\imath}j,k}\chi^j_-\bar{\chi}^k_+ \: - \:
\bar{\mathcal{W}}_{\bar{\imath}}\right)
\: - \: i \bar{\mathcal{W}}_{\bar{\imath}\bar{\jmath}}\bar{\chi}^{\bar{\imath}}_-\chi^{\bar{\jmath}}_+ \\
&\: + \: g_{i\bar{\jmath},k\bar{l}}\bar{\chi}_+^i
\chi_+^{\bar{\jmath}} \chi_-^k \bar{\chi}_-^{\bar{l}} \\
& \: + \: \frac{i}{r} \mathcal{W} \: - \: \frac{i}{r} \bar{\mathcal{W}}
,
\end{split}
\end{equation}
with, assuming $Y^i$ is chosen holomorphic, supersymmetry transformations
\begin{equation}
\begin{split}
\delta \rho^i &= i \bar{\zeta}_+\chi^i_- + i \zeta_-\bar{\chi}^i_+,\\
\delta \bar{\rho}^{\bar{\imath}} &= i \bar{\zeta}_- \chi^{\bar{\imath}}_+
 + i \zeta_+ \bar{\chi}^{\bar{\imath}}_-,\\
\delta \bar{\chi}^i_+ &= -2 \bar{\zeta}_-\bar{\partial}\rho^i
- \bar{\zeta}_+ \left( G^i + \frac{i}{r} Y^i \right),\\
\delta \chi^i_- &= -2 \zeta_+\partial\rho^i +\zeta_- \left( G^i
+ \frac{i}{r} Y^i \right),\\
\delta \chi^{\bar{\imath}}_+ &= -2 \zeta_-\bar{\partial}
\bar{\rho}^{\bar{\imath}} - \zeta_+ \left( \bar{G}^{\bar{\imath}} -
\frac{i}{r} Y^{\bar{\imath}} \right),\\
\delta \bar{\chi}^{\bar{\imath}}_- &= -2 \bar{\zeta}_+
\partial\bar{\rho}^{\bar{\imath}}  +\bar{\zeta}_- \left(
\bar{G}^{\bar{\imath}} - \frac{i}{r} Y^{\bar{\imath}} \right),\\
\delta G^i &= 2i(\zeta_+\tilde{\nabla}_z\bar{\chi}^i_+ - \bar{\zeta}_-
\tilde{\nabla}_{\bar{z}}\chi^i_-) - \frac{i}{r} \left( i \bar{\zeta}_+ \chi_-^j
+ i \zeta_- \bar{\chi}_+^j \right) \partial_j Y^i,\\
\delta \bar{G}^{\bar{\imath}} &= 2i(\bar{\zeta}_+
\tilde{\nabla}_z\chi^{\bar{\imath}}_+
- \zeta_-\tilde{\nabla}_{\bar{z}}\bar{\chi}^{\bar{\imath}}_-)
+ \frac{i}{r} \left( i \bar{\zeta}_- \chi_+^{\bar{\jmath}} + i
\zeta_+ \bar{\chi}_-^{\bar{\jmath}} \right) \partial_{\bar{\jmath}}
Y^{\bar{\imath}},
\end{split}
\end{equation}
provided that the Killing spinor equations~(\ref{eq:killing-twisted})
% and the
%constraint~(\ref{constraint1})
are satisfied.
To recover \cite{gomis-lee}[equ'ns (4.2), (4.5)] on flat target spaces,
take $Y^i = - i \Delta \rho^i$.

\section{Topological twists}  \label{sect:tft}

When the superpotential vanishes (and, for ordinary chiral multiplets,
$X = 0$), there are no curvature couplings, and the theories admit the
same topological twists discussed in {\it e.g.} \cite{witten-tft}.

When the superpotential is nonzero, this story is more complicated.
Sometimes those curvature couplings are incompatible with the topological
twist (which can sometimes be alleviated by twisting bosons as in
{\it e.g.} \cite{alg22,alg02}).  However, a more fundamental problem
is that even if one can consistently twist the theory,
the curvature terms induced by the superpotential break the
BRST symmetry, in the sense that the action is no longer BRST closed.
Those terms are compatible with supersymmetry only so long
as the supersymmetry parameters $\zeta$, $\bar{\zeta}$ obey the Killing
spinor equation, and serve to `mop up' the curvature-dependent terms
that result from using the Killing spinor equations.  Since in a topological
field theory the BRST transformations are parametrized by a scalar,
the Killing spinor equations are not relevant, and so the
curvature-dependent terms in the action are extraneous, breaking the
BRST symmetry.

However, just because we cannot always twist, does not imply that topological
field theories do not exist.  For example, the papers \cite{alg22,alg02}
describe examples of two-dimensional topological field theories
with nonzero superpotential, on two-spheres and other two-dimensional
worldsheets.  These topological field theories were {\it not} obtained
by topological twisting of a two-dimensional theory with curvature
couplings.  Instead, they were obtained by twisting a flat-worldsheet
theory.  The result continues to make sense on two-spheres and other
two-dimensional worldsheets because the BRST transformations are parametrized
by a scalar; the action neither needs nor contains curvature-dependent
terms.

Thus, to summarize, if the superpotential vanishes (and $X=0$), then
topological twistings exist.  If the superpotential is nonzero, there still
exist topological field theories, but they are not obtained by twisting
an action that includes curvature-dependent terms.

\section{Discussion}   \label{sect:disc}

In this note we have discussed curvature couplings in general nonlinear
sigma models with potential, for both ordinary and twisted chiral multiplets,
following a program initiated in \cite{allanetal,fest-seib}.
In the special case of target spaces that are vector spaces, such couplings
have been previously discussed in {\it e.g.}
\cite{benini-cremonesi,doroud-gomis-le floch-lee}; our purpose here was to
generalize such couplings to general target spaces (and with $U(1)_R$ actions
described by general holomorphic Killing vectors), so as to give
some degree of insight into the structure of their actions.  We have
also discussed
general issues surrounding existence of topological twists in this
context.

It would be interested to understand localization
in these more general theories, and {\it e.g.} compare the results of
localization in A-twisted Landau-Ginzburg models of the form considered in
\cite{alg22} to corresponding Landau-Ginzburg models with curvature interactions
as described here.  The A-twisted Landau-Ginzburg models realized
Mathai-Quillen classes which gave a mathematical understanding of the
behavior of renormalization group flow, as a type of Thom class
phenomenon.  (See also \cite{ando-sharpe} for closely
analogous behaviors in elliptic
genera.)  It would be interesting to see if
something analogous arises in localization computations.

\section{Acknowledgements}

We thank Robert Bryant, Josh Lapan,
Ilarion Melnikov, Daniel Park, and Ronen Plesser for helpful conversations.
This work was partially supported by NSF grant PHY-1068725.

\appendix

\section{Twisted chiral superfields without curvature couplings}
\label{app:twistedchirals}

As twisted chiral multiplets are discussed less commonly than ordinary
chiral multiplets, it will be useful to review a few of their
basic properties.  (See also \cite{hori-ms}[section 12.2] for another
comparison of ordinary and twisted chiral multiplets.)
Given twisted chiral multiplets of the form
$(\rho^i, \bar{\chi}_+^i, \chi_-^i)$, where the $\rho$ are bosons,
the Lagrangian for a theory containing
only twisted chiral multiplets on $\mathbb{R}^2$ is of the form
\begin{equation}
\begin{split}
\mathcal{L}_T =
& g_{i\bar{\jmath}}\partial_{m}\rho^i\partial^{m}\bar{\rho}^{
\bar{\jmath}}  +2ig_{i\bar{\jmath}}\bar{\chi}^{\bar{\jmath}}_{-}\nabla_{\bar{z}}\chi^i_{-}
+ 2ig_{i\bar{\jmath}}\chi^{\bar{\jmath}}_{+}\nabla_{z}\bar{\chi}^i_{+}
+  g_{i\bar{\jmath}}G^i\bar{G}^{\bar{\jmath}}\\
& - G^i\left(g_{i\bar{\jmath},\bar{k}} \bar{\chi}^{\bar{\jmath}}_-
\chi^{\bar{k}}_+ - \mathcal{W}_i\right) - i \mathcal{W}_{ij}
\chi^i_-\bar{\chi}^j_+ \\
& -  \bar{G}^{\bar{\imath}}
 \left(g_{\bar{\imath}j,k}\chi^j_-\bar{\chi}^k_+ -
\bar{\mathcal{W}}_{\bar{\imath}}\right)
 - i \bar{\mathcal{W}}_{\bar{\imath}\bar{\jmath}}\bar{\chi}^{\bar{\imath}}_-\chi^{\bar{\jmath}}_+ \\
&- g_{i\bar{\jmath},k\bar{l}}\bar{\chi}_+^i
\chi_+^{\bar{\jmath}} \chi_-^k \bar{\chi}_-^{\bar{l}}
,
\end{split}
\end{equation}
with supersymmetry transformations
\begin{equation}
\begin{split}
\delta \rho^i &= i \bar{\zeta}_+\chi^i_- + i \zeta_-\bar{\chi}^i_+,\\
\delta \bar{\rho}^{\bar{\imath}} &= i \bar{\zeta}_- \chi^{\bar{\imath}}_+
 + i \zeta_+ \bar{\chi}^{\bar{\imath}}_-,\\
\delta \bar{\chi}^i_+ &= -2 \bar{\zeta}_-\bar{\partial}\rho^i
- \bar{\zeta}_+ G^i,\\
\delta \chi^i_- &= -2 \zeta_+\partial\rho^i +\zeta_- G^i,\\
\delta \chi^{\bar{\imath}}_+ &= -2 \zeta_-\bar{\partial}
\bar{\rho}^{\bar{\imath}} - \zeta_+ \bar{G}^{\bar{\imath}},\\
\delta \bar{\chi}^{\bar{\imath}}_- &= -2 \bar{\zeta}_+
\partial\bar{\rho}^{\bar{\imath}}  +\bar{\zeta}_- \bar{G}^{\bar{\imath}},\\
\delta G^i &= 2i(\zeta_+\tilde{\nabla}_z\bar{\chi}^i_+ - \bar{\zeta}_-
\tilde{\nabla}_{\bar{z}}\chi^i_-) ,\\
\delta \bar{G}^{\bar{\imath}} &= 2i(\bar{\zeta}_+
\tilde{\nabla}_z\chi^{\bar{\imath}}_+
- \zeta_-\tilde{\nabla}_{\bar{z}}\bar{\chi}^{\bar{\imath}}_-) .
\end{split}
\end{equation}
They can be obtained by taking the limit of $r\rightarrow \infty$
in the corresponding theory on $S^2$ discussed in
section~\ref{sect:twistedchiral}.
If we integrate out the auxiliary fields $G^i$, $\bar{G}^i$, the lagrangian
takes the simpler form
\begin{equation}
\begin{split}
\mathcal{L}_T =
& g_{i\bar{\jmath}}\partial_{m}\rho^i\partial^{m}\bar{\rho}^{
\bar{\jmath}}  +2ig_{i\bar{\jmath}}\bar{\chi}^{\bar{\jmath}}_{-}\nabla_{\bar{z}}\chi^i_{-}
+ 2ig_{i\bar{\jmath}}\chi^{\bar{\jmath}}_{+}\nabla_{z}\bar{\chi}^i_{+}
+ R_{i \bar{\jmath} k \bar{\ell}} \bar{\chi}^i_+ \chi^{\bar{\jmath}}_+
\chi^k_- \bar{\chi}^{\bar{\ell}}_-
\\
& - g^{i \bar{\jmath}} \mathcal{W}_i \bar{\mathcal{W}}_{\bar{\jmath}}
-i \chi_-^i \bar{\chi}^j_+ D_i \partial j \mathcal{W}
-i \bar{\chi}^{\bar{\imath}} \chi^{\bar{\jmath}}_+
D_{\bar{\imath}} \partial_{\bar{\jmath}} \bar{\mathcal{W}}
\end{split}
\end{equation}
with supersymmetry transformations
\begin{equation}
\begin{split}
\delta \rho^i &= i \bar{\zeta}_+\chi^i_- + i \zeta_-\bar{\chi}^i_+,\\
\delta \bar{\rho}^{\bar{\imath}} &= i \bar{\zeta}_- \chi^{\bar{\imath}}_+
+ i \zeta_+ \bar{\chi}^{\bar{\imath}}_-,\\
\delta \bar{\chi}^i_+ &= -2 \bar{\zeta}_-\bar{\partial}\rho^i
+ \bar{\zeta}_+ \left( \Gamma^i_{jk} \chi_-^j \bar{\chi}_+^k
+ g^{i \bar{\jmath}} \bar{\mathcal{W}}_{\bar{\jmath}} \right),\\
\delta \chi^i_- &= -2 \zeta_+\partial\rho^i +
\zeta_- \left( \Gamma^i_{jk} \chi_-^j \bar{\chi}_+^k
+ g^{i \bar{\jmath}} \bar{\mathcal{W}}_{\bar{\jmath}} \right),\\
\delta \chi^{\bar{\imath}}_+ &= -2\zeta_-\bar{\partial}\bar{\rho}^{\bar{\imath}}
+ \zeta_+ \left( \Gamma^{\bar{\imath}}_{\bar{\jmath} \bar{k}}
\bar{\chi}^{\bar{\jmath}}_- \chi^{\bar{k}}_+ +
g^{\bar{\imath} j} \mathcal{W}_j \right), \\
\delta \bar{\chi}^{\bar{\imath}}_- &= -2 \bar{\zeta}_+\partial\bar{\rho}^{
\bar{\imath}} + \bar{\zeta}_- \left(
\Gamma^{\bar{\imath}}_{\bar{\jmath} \bar{k}}
\bar{\chi}^{\bar{\jmath}}_- \chi^{\bar{k}}_+ +
g^{\bar{\imath} j} \mathcal{W}_j \right).
\end{split}
\end{equation}
For contrast,
compare the Lagrangian for ordinary chiral multiplets $(\phi^i, \psi_+^i,
\psi_-^i)$ given in
\cite{witten-tft}[equ'n (2.4)].  Modulo signs and irrelevant factors, the
lagrangian for a theory of purely twisted chiral multiplets is the
same as the action for a theory of purely ordinary chiral multiplets -- the
difference between the two lies in the supersymmetry transformations.

Note that if one were to dualize the tangent bundle $TM$ to $T^*M$ on
the right-movers, the effect would be to convert the twisted chiral
multiplets back into chiral multiplets.  This is a special case of a
duality in (0,2) theories discussed in {\it e.g.} \cite{es02b}.

Now, let us specialize to the case $\mathcal{W} = 0$, in which case there
are no curvature couplings on any worldsheet,
and consider topological twists.  It is straightforward to topologically
twist the theory of twisted chiral multiplets; however, the results are
identical to the topological field theories obtained from twisting theories
of ordinary chiral multiplets.  Specifically, the A-twist of twisted chiral
multiplets is equivalent to the B-twist of ordinary chiral multiplets,
and conversely, as one would expect from ideas related to aspects of
(2,2) algebras in two dimensions.

It will be useful to make that correspondence explicit.
Adapting the conventions of
\cite{witten-tft}, the A twist is defined by making the supersymmetry
transformation parameters $\zeta_+$, $\bar{\zeta}_-$ into (Grassmann-valued)
scalars, parametrizing the BRST transformations.  The same twist here would
result in fermions coupling to bundles as
\begin{displaymath}
\begin{array}{ll}
\bar{\chi}_+^i \: \in \: \Gamma_{C^{\infty}}\left( K_{\Sigma}
\otimes \rho^* T^{1,0} \tilde{M}
\right), &
\chi_-^i \: \in \: \Gamma_{C^{\infty}}\left(
\overline{K}_{\Sigma} \otimes \left( \rho^*  T^{0,1} \tilde{M}
\right)^* \right), \\
\chi_+^{\bar{\imath}} \: \in \: \Gamma_{C^{\infty}}\left(
\left( \rho^* T^{1,0} \tilde{M} \right)^*
\right), &
\bar{\chi}_-^{\bar{\imath}} \: \in \: \Gamma_{C^{\infty}}\left(
\rho^* T^{0,1} \tilde{M} \right).
\end{array}
\end{displaymath}
where $\Sigma$ is the worldsheet, $K_{\Sigma}$ and
$\overline{K}_{\Sigma}$ are the holomorphic and antiholomorphic canonical
bundles,
and BRST transformations
\begin{eqnarray*}
\delta \rho^i & = & 0, \\
\delta \rho^{\bar{\imath}} & = &
i \bar{\zeta}_- \chi^{\bar{\imath}}_+  + i
\zeta_+ \bar{\chi}^{\overline{\imath}}_-,
\\
\delta \bar{\chi}^i_+ & = & -2  \bar{\zeta}_- \bar{\partial} \rho^i, \\
\delta \chi^i_- & = & -2  \zeta_+ \partial \rho^i, \\
\delta \chi^{\bar{\imath}}_+ & = & \zeta_+ \Gamma^{\bar{\imath}}_{
\bar{\jmath} \bar{k}} \bar{\chi}^{\bar{\jmath}}_- \chi^{\bar{k}}_+, \\
\delta \bar{\chi}^{\bar{\imath}}_- & = & \bar{\zeta}_- \Gamma^{\bar{\imath}}_{
\bar{\jmath} \bar{k}} \bar{\chi}^{\bar{\jmath}}_- \chi^{\bar{k}}_+ .
\\
\end{eqnarray*}
This structure, the A-twisted theory of twisted chiral multiplets,
can easily be seen to be equivalent to the B-twisted theory of
ordinary chiral multiplets described in \cite{witten-tft}, and as such
is only well-defined in the special case that $K_{\tilde{M}}^2$ is
trivial \cite{es02b}.

For completeness, let us examine the opposite case.
In the B-twisted theory of ordinary chiral multiplets described in
\cite{witten-tft}, the supersymmetry transformation parameters
$\bar{\zeta}_+$, $\bar{\zeta}_-$ are (Grassmann-valued) scalars.
This twist results in fermions coupling to bundles as
\begin{displaymath}
\begin{array}{ll}
\bar{\chi}_+^i \: \in \: \Gamma_{C^{\infty}}\left( K_{\Sigma}
\otimes \rho^* T^{1,0} \tilde{M}
\right), &
\chi_-^i \: \in \: \Gamma_{C^{\infty}}\left( \left( \rho^*  T^{0,1} \tilde{M}
\right)^* \right), \\
\chi_+^{\bar{\imath}} \: \in \: \Gamma_{C^{\infty}}\left(\left( \rho^* T^{1,0} \tilde{M} \right)^*
\right), &
\bar{\chi}_-^{\bar{\imath}} \: \in \: \Gamma_{C^{\infty}}\left(
\overline{K}_{\Sigma} \otimes \rho^* T^{0,1} \tilde{M} \right).
\end{array}
\end{displaymath}
and the BRST transformations are given by
\begin{eqnarray*}
\delta \rho^i & = & i \bar{\zeta}_+ \chi^i_-, \\
\delta \bar{\rho}^{\bar{\imath}} & = &
i \bar{\zeta}_- \chi^{\bar{\imath}}_+, \\
\delta \bar{\chi}^i_+ & = & -2  \bar{\zeta}_- \bar{\partial}
\rho^i + \bar{\zeta}_+ \Gamma^i_{jk} \chi^j_- \bar{\chi}^k_+, \\
\delta \chi^i_- & = & 0, \\
\delta \chi^{\bar{\imath}}_+ & = & 0, \\
\delta \bar{\chi}^{\bar{\imath}}_- & = &
-2  \bar{\zeta}_+ \partial \bar{\rho}^i +
\bar{\zeta}_- \Gamma^{\bar{\imath}}_{\bar{\jmath} \bar{k}}
\bar{\chi}^{\bar{\jmath}}_- \chi^{\bar{k}}_+ . \\
\end{eqnarray*}
This structure, the B-twisted theory of twisted chiral multiplets,
can easily be seen to be equivalent to the A-twisted theory of
ordinary chiral multiplets described in \cite{witten-tft}.

\end{document}